\documentclass[11pt]{article}

\usepackage{amsthm,amsmath,amssymb,amsfonts,xspace,graphics,enumerate}
\usepackage{color}
\usepackage{authblk}
\usepackage{cite}

\usepackage{graphicx}
\usepackage{array,rotating}
\usepackage{amsmath}
\usepackage{amsfonts}
\usepackage{lscape}
\usepackage{natbib}
\usepackage{multirow}
\usepackage{booktabs}
\usepackage{mdwlist}
\usepackage{enumitem} 
\setlist[itemize]{noitemsep, topsep=0.5pt}

\usepackage{bm} 
\usepackage{mathtools} 
\usepackage[table]{xcolor}
\usepackage{url} 
\usepackage{wrapfig}

\usepackage{soul}

\usepackage{float}

\usepackage{natbib}
\usepackage{comment}
\usepackage{fullpage} 
\usepackage{graphicx}
\usepackage[font=small]{caption}
\usepackage{subcaption}
\usepackage{amssymb}
\usepackage{amsmath}
\usepackage{graphics}
\usepackage{bm}
\usepackage{color}
\usepackage{amsthm}
\usepackage{enumerate}
\usepackage{amsfonts}
\usepackage{bbm}
\usepackage[toc]{appendix}
\usepackage{enumitem}
\usepackage{xr}
\usepackage{changes}

\usepackage[compact]{titlesec}
\titlespacing{\section}{0pt}{2ex}{1ex}
\titlespacing{\subsection}{0pt}{1ex}{0ex}
\titlespacing{\subsubsection}{0pt}{0.5ex}{0ex}

\usepackage{tikz}
\usepackage[CJKbookmarks=true,
bookmarksnumbered=true,
bookmarksopen=true,
colorlinks=true,
citecolor=blue,
linkcolor=blue,
anchorcolor=blue,
urlcolor=blue]{hyperref}
\usepackage[ruled, vlined, lined, commentsnumbered,linesnumbered]{algorithm2e}

\newcommand{\bzero}{{\bf 0}}

\newcommand{\bA}{{\bf A}}

\newcommand{\bC}{{\bf C}}

\newcommand{\bW}{{\bf W}}

\newcommand{\hp}{{\hat{p}}}

\newcommand{\bh}{{\bf h}}

\newcommand{\bu}{{\bf u}}

\newcommand{\bw}{{\bf w}}

\newcommand{\alpham}{\mbox{\boldmath $\alpha$}}
\newcommand{\gammam}{\mbox{\boldmath $\gamma$}}

\newcommand{\halpha}{{\hat{\alpha}}}

\newcommand{\hbe}{{\hat{\beta}}}

\newcommand{\halpham}{{\hat{\alpham}}}

\newcommand{\htau}{{\hat{\tau}}}

\newcommand{\iid}{i.i.d.\xspace}

\newtheorem{ass}{Assumption}

\newcommand{\tGFPLM}{{\textsf{GFPLM }}}
\newcommand{\tFGAM}{{\textsf{FGAM }}}
\newcommand{\tCBPS}{{\textsf{CBPS}}}
\newcommand{\tKBCB}{{\textsf{KBCB }}}

\newtheorem{rmk}{Remark}

\begin{document}

\title{\bf Average Treatment Effect Estimation in Observational Studies with Functional Covariates}
\author{Rui Miao} 
\author{Wu Xue} 
\author{Xiaoke Zhang} 
\affil{George Washington University}
\maketitle

\bigskip
\begin{abstract}
Functional data analysis is an important area in modern statistics and has been successfully applied in many fields. 
Although many scientific studies aim to find causations, a predominant majority of functional data analysis approaches can only reveal correlations. In this paper, average treatment effect estimation is studied for observational data with functional covariates.
This paper generalizes various state-of-art propensity score estimation methods for multivariate data to functional data. 
The resulting average treatment effect estimators via propensity score weighting
are numerically evaluated by a simulation study and applied to a real-world dataset to study the causal effect of duloxitine on the pain relief of chronic knee osteoarthritis patients.
\end{abstract}







\section{Introduction} \label{sec:intro}

Functional data analysis (FDA) has become increasingly important in modern statistics and has been successfully applied in a variety of scientific fields. 
Apart from books on general introductions to FDA \citep[e.g.,][]{Bosq00, RamsS05, FerrV06, HorvK12, 
HsinE15, KokoR17}, 
recent advances of FDA, 
including innovative methodologies, profound theories, efficient algorithms, and successful applications,
have been illustrated by numerous survey papers \citep[e.g.,][]{Guo04, Muel08, DeliGC10, Geen11, HormK12, Cuev14, MarrA14, Shan14, WangCM16, ChenZP17, Nagy17, Vieu18, KokoR19}.

A majority of FDA methods can only reveal correlations primarily via either functional regression models \citep[for reviews see e.g.,][]{Morr15, GrevS17, PagaS17, ReisGS17} or correlation measures \citep[e.g.,][]{LeurMS93, DubiM05, CupiEG08, EubaH08, Lian14, ShinL15, ZhouLW18}. However, FDA methods for causal inference is underdeveloped despite the importance of causation in many scientific studies. 
Among very few exceptions, almost all of them focus on randomized clinical trials \citep[e.g.,][]{Lind12, McKeQ14, CiarPO15, CiarPO18, ZhaoLL18, ZhaoL19}. 
In classical causal inference for observational studies where multivariate data are of primary interest, the propensity score \citep{RoseR83} plays an important role and has been widely applied in epidemiology and political science among others. Despite its popularity, its use in FDA to study causations in observational studies 
is nearly void. 

The main contribution of this paper is to introduce and adapt various state-of-art propensity score methods to observational functional data. We consider the scenario where the treatment is binary and at least one covariate is functional. We generalize the definition of the propensity score to functional data, and study two types of propensity score estimations. The propensity score is estimated by either directly fitting a functional regression model or balancing appropriate functions of the covariates.
This paper in particular focuses on propensity score weighting \citep[e.g.,][]{Rose87, RobiHB00, HiraIR03}, although the propensity score may be used to adjust for confounding through other means, e.g., matching \citep[e.g.,][]{RoseR85, Rose89, AbadI06} and subclassification \citep[e.g.,][]{RoseR84, Rose91, Hans04}.  A systematic comparison of two popular propensity-score-weighted average treatment effect estimators is provided in both a simulation study and a real data application.

The rest of the paper proceeds as follows. Section \ref{sec:frame} provides the problem setup and generalizes the classical definition of the propensity score 
to functional data where the treatment is binary and one covariate is functional.
Section \ref{sec:method} introduces two types of propensity score estimations via direct modeling and covariate balancing respectively and two widely used average treatment effect estimators via propensity score weighting. The two average treatment effect estimators based on a variety of estimated propensity score weights are comprehensively 
compared in a simulation study in Section \ref{sec:simu}. They are also applied in a real data analysis in Section \ref{sec:data} to study the causal effect of duloxitine on the pain relief of chronic knee osteoarthritis patients. Discussion in Section \ref{sec:disc} concludes the paper.

\section{Framework} \label{sec:frame}





Suppose that $Y$ is a continuous outcome, $T$ is a binary treatment variable which equals either $0$ (control) or $1$ (treatment), $\bW$ is a multivariate covariate, and $X(\cdot)$ is a functional covariate defined on a compact domain $\mathcal{T}$. Suppose that $E \int_\mathcal{T}\{X(t)\}^2\, dt < \infty$ and $X(\cdot)$ is smooth, e.g., continuous or twice-differentiable. 
Without loss of generality $E(\bW)=\bzero$, $\mathcal{T}=[0, 1]$ and $E\{X(t)\}=0$ for all $t \in [0,1]$.

Let $Y(1)$ and $Y (0)$ represent the potential values of $Y$ when $T=1$ and $0$ respectively. In practice $Y=T Y(1) + (1-T) Y(0)$ is observable but $Y(1)$ and $Y (0)$ are not both observable.
Based on $\{(Y_i, T_i, \bW_i, \{X_i(t): t \in [0, 1]\}): i=1,\ldots, n\}$, which are independently and identically distributed (\iid) copies of $(Y, T, \bW, \{X(t): t \in [0, 1]\})$, we aim to estimate the average treatment effect $\tau=E\{Y(1)-Y(0)\}$.

We assume that each $X_i(\cdot)$ is fully observed, but the methods below are also applicable for densely measured $X_i(\cdot)$ since its entire trajectory can be accurately recovered by smoothing \citep[e.g.,][]{ZhanC07}. In this paper we only consider low-dimensional $\bW_i$. The handling of high-dimensional multivariate covariates is beyond the scope of this paper but is a promising topic for future research \citep[e.g.,][]{BellCH14, Farr15, BellCF17, CherCD18, NingPI18}.

Similar to its classical counterpart \citep[e.g.,][]{RoseR83, RobiHB00}, the propensity score is defined by 
\begin{equation}\label{eq:ps}
p(\bW_i, X_i) = P(T_i = 1 \mid \bW_i, X_i(\cdot)).
\end{equation}

In this paper we make the following two assumptions:
\begin{ass}
\label{ass:ignore}
$$
T_i \perp \left(Y_i(0), Y_i(1) \right) \mid \left(\bW_i, X_i(\cdot)\right), 
$$
where 
``$\perp$'' represents independence. 
\end{ass}

\begin{ass}
\label{ass:positive}
The propensity score satisfies
$$
0 < P(T_i = 1 \mid \bW_i = \bw, X_i(\cdot) = x(\cdot)) < 1,
$$
for all vectors $\bw$ and all functions $x(\cdot)$ defined on $[0, 1]$ such that $\int_0^1 \{x(t)\}^2\, dt < \infty$ . 
\end{ass}
Assumptions \ref{ass:ignore} and \ref{ass:positive} are straightforward generalizations of the commonly used \emph{strong ignorability} and \emph{positivity} assumptions in classical causal inference respectively. Assumption \ref{ass:ignore} 
implies that there is no unmeasured covariate, while Assumption \ref{ass:positive} essentially requires
that every sample has a positive probability of 
receiving the treatment or being in the control group.

\section{Methodology}\label{sec:method}
In this section, we introduce various methods for propensity score estimation and two average treatment effect estimators via propensity score weighting. 

\subsection{Propensity Score Estimation}\label{sec:pse}
We propose two types of propensity score estimations, one via direct modeling and the other via covariate balancing. They will be introduced in Sections \ref{sec:dm} and \ref{sec:cb} respectively.
\subsubsection{Direct Modeling}\label{sec:dm}
To estimate the propensity score $p(\bW_i, X_i)$, one may assume a parametric model for $p(\bW_i, X_i)$ and fit it with an appropriate estimation procedure. 

The simplest model might be the generalized functional partial linear model (GFPLM): 
\begin{equation}\label{eq:gfplm}
\text{logit}\left\{p(\bW_i, X_i) \right\} = \alpha_0+ \alpham_1^\top \bW_i + \int_0^1 \beta(t) X_i(t)\, dt,
\end{equation}
where $\text{logit}(x) = \log\{x / (1-x)\}$ and the scalar $\alpha_0$, vector $\alpham_1$ and function $\beta(\cdot)$ are unknown parameters to be estimated.
Apparently the GFPLM 
is an extension of \cite{Jame02}, \cite{MuelS05} and \cite{Shin09}. 


{Similar to fitting other functional regression models considered in the FDA literature, regularization for the functional coefficient $\beta$ is needed to fit (\ref{eq:gfplm}). Popular regularizations include the truncated basis function expansion \citep[e.g., ][Ch.4]{CardFS99, RamsS05}, roughness penalization \citep[e.g., ][]{YuanC10} and their combinations \citep[e.g., ][Ch.5]{CardFS03, RamsS05}. The 
most straightforward regularization is perhaps the first one above 
where the basis functions are obtained by functional principal component analysis (FPCA). Explicitly, 
}
the functional covariate may be approximated by by $X_i(t) \approx \sum_{k=1}^L A_{ik} \phi_k(t)$ where $\phi_k(\cdot), 1 \leq k \leq L < \infty,$ are eigenfunctions corresponding to the top $L$ eigenvalues $\lambda_1 \geq \cdots \geq \lambda_L > 0$ of the covariance function $\text{Cov}\{X(s), X(t)\}$, and $A_{ik}=\int_0^1 X_i(t) \phi_k(t), 1 \leq k \leq L,$ are corresponding FPC scores. Thus 
\begin{equation}\label{eq:agfplm}
\text{logit}\left\{p(\bW_i, X_i) \right\} \approx \alpha_0+ \alpham_1^\top \bW_i + \sum_{k=1}^L \beta_k A_{ik},
\end{equation}
where $\beta_k= \int_0^1 \beta(t) \phi_k(t)\, dt, 1 \leq k \leq L$.
The maximum likelihood method can be used to find the parameter estimates $(\halpha_0, \halpham_1, \hbe_1, \ldots, \hbe_L)$ and thus the propensity score estimate $\hp(\bW_i, X_i)$. 

\begin{rmk}\label{rmk:gfplm}
\item 1. 
The terms $\phi_k(\cdot), \lambda_k, A_{ik}, k \geq 1$ above are all population quantities. In practice one can only obtain their sample versions.
\item 2. The aforementioned FPCA-regularized maximum likelihood method is also applicable to fit a GPFLM if $X_i$ is 
a multidimensional functional covariate, i.e., 
$$
\text{logit}\left\{p(\bW_i, X_i) \right\} = \alpha_0+ \alpham_1^\top \bW_i + \int \beta(\bu) X_i(\bu)\, d \bu,
$$
where $\bu$ is a generic and multidimensional index for 
$X_i$. With the FPC scores obtained by FPCA, this multidimensional GPFLM can also be approximated by (\ref{eq:agfplm}) and fitted by the maximum likelihood method.
\item {3. Following the suggestion by \cite{Rubi07} that the propensity score estimation is conducted without access to outcome data, 
the number of FPC scores 
$L$, a tuning parameter, can be determined by various means when estimating the propensity score, including the fraction of variation explained (e.g., 95\% or 99\%), cross-validation, the Akaike information criterion (AIC), etc.}
\end{rmk}

In addition to the GFPLM, 
one may fit other propensity score models, such as the functional generalized additive model 
\citep[FGAM, e.g.,][]{MuelWY13, McLeHS14}: 
\begin{equation}\label{eq:fgam}
\text{logit}\left\{p(\bW_i, X_i) \right\} = \alpha_0 + \alpham_1^\top \bW_i + \int_0^1 \eta(t, X_i(t))\, dt, 
\end{equation}
where the unknown parameters are $\alpha_0$, a scalar, $\alpham_1$, a vector, and $\eta(*, \cdot)$, a bivariate function. To fit 
(\ref{eq:fgam}), one may apply the maximum likelihood method after approximating $\eta(*, \cdot)$ by a set of tensor products of B-spline basis functions. 

\subsubsection{Covariate Balancing}\label{sec:cb}


In the classical literature on causal inference, it is well known that parametric methods for propensity score estimation may suffer from model misspecification substantially \citep[e.g.,][]{SmitT05, KangS07}. 
Recently covariate balancing methods, which aim to mimic randomization, have been proposed as important alternatives 
\citep[e.g.,][]{QinZ07, Hain12, ImaiR14, Zubi15, LiMZ18, WongC18, Zhao19}. 
To the best of our knowledge, all existing covariate balancing methods so far are developed to handle multivariate covariates and cannot be directly used for functional covariates. 


{One way of balancing both multivariate covariates $\bW_i$ and functional covariate $X_i$ is to generalize the covariate balancing propensity score (CBPS) method \citep{ImaiR14} by considering the following functional covariate balancing equation:
\begin{equation} \label{eq:fcb}
\frac{1}{n} \sum_{i=1}^n \left\{ \frac{T_i}{p(\bW_i, X_i)} - \frac{(1-T_i)}{1-p(\bW_i, X_i)} \right\} \bh(\bW_i, X_i) = \bzero, 
\end{equation}
where a parametric model is assumed for the propensity score $p(\bW_i, X_i)$ and $\bh(\bW_i, X_i)$ is a user-defined vector-valued function to reflect how 
$\bW_i$ and 
$X_i$ are balanced.} 

{To solve (\ref{eq:fcb}),} 
we propose to substitute the functional covariate $X_i$ by a multivariate covariate.
For example, by FPCA as in Section \ref{sec:dm}, the functional covariate $X_i(\cdot)$ can be approximated by $X_i(t) \approx \sum_{k=1}^L A_{ik} \phi_k(t)$ with a proper integer $L$ such that $\bA_i=(A_{i1}, \ldots, A_{iL})^\top$ possesses a majority of the information of $X_i$. Thus $\bA_i$ may be used as a substitute of $X_i$ and we {may replace $p(\bW_i, X_i)$ and $\bh(\bW_i, X_i)$ in (\ref{eq:fcb}) by}
the \emph{substitute propensity score} 
\begin{equation}\label{eq:aps}
p(\bC_i) = P(T_i = 1 \mid \bC_i),
\end{equation}
{and another user-defined vector-valued function $\bh(\bC_i)$ respectively,} 
where $\bC_i = (\bW_i^\top, \bA_i^\top)^\top$. Obviously if $X(\cdot)$ is of finite rank in terms of its spectral decomposition, i.e., FPCA, the substitute propensity score $p(\bC_i)$ is equivalent to $p(\bW_i, X_i)$
when $L$ is chosen to be the rank of $X(\cdot)$. {Then it suffices to solve 
\begin{equation}\label{eq:mcb}
\frac{1}{n} \sum_{i=1}^n \left\{ \frac{T_i }{p(\bC_i)} - \frac{(1-T_i)}{1-p(\bC_i)}  \right\} \bh(\bC_i)= \bzero,
\end{equation}
which is exactly the covariate balancing equation for multivariate covariates \citep{ImaiR14}. }

To solve (\ref{eq:mcb}), 
one may assume a logistic model for $p(\bC_i)$: 
\begin{equation}\label{eq:gfplmc}
\text{logit}\left\{p(\bC_i) \right\} = \gamma_0 + \gammam_1^\top \bC_i,
\end{equation}
where $\gamma_0$ and $\gammam_1$ are unknown parameters to be estimated, which is equivalent to the approximate GFPLM in (\ref{eq:agfplm}). {One also needs to specify $\bh(\bC_i)$ to reflect how $\bC_i$ are balanced. }
For example, one may define $\bh(\bC_i) = \bC_i$ to balance the first moment of $\bC_i$. Alternatively to balance both the first and second moments of $\bC_i$, one may use $\bh(\bC_i) = (\bC_i^\top, (\bC_i^2)^\top)^\top$ where $\bC_i^2$ contains the entry-wise square of $\bC_i$. {Then we let $\hp(\bW_i, X_i)=\hp(\bC_i)$ obtained in (\ref{eq:mcb}).}

\subsection{Average Treatment Effect Estimation}\label{sec:ate}




To estimate the average treatment effect $\tau=E\{Y(1)-Y(0)\}$, a variety of estimators via propensity score weighting have been proposed, such as the Horvitz-Thompson estimator \citep{HorvT52}, the inverse propensity score weighting estimator \citep{HiraIR03}, the weighted least squares regression estimator \citep{RobiHB00, FreeB08}, and the doubly robust estimator \citep{RobRZ94} among others.

In the simulation experiments and real data application below, we will consider two representative average treatment effect estimators, the Horvitz-Thompson (HT) estimator and 
H\'{a}jek estimator, to numerically evaluate and compare the propensity score estimation methods in Section \ref{sec:pse}.

Explicitly, for each propensity score estimate $\hp_i=\hp(\bW_i, X_i)$ obtained by either direct modeling or covariate balancing approach, the HT and H\'{a}jek estimators are respectively defined by
\begin{eqnarray*}
\htau_\text{HT} &=& \frac{1}{n}\sum_{i=1}^n \left\{\frac{T_i Y_i}{\hp_i} - \frac{(1-T_i) Y_i}{1-\hp_i} \right\}, \quad \text{and}\\
\htau_\text{H\'{a}jek} &=& \frac{\sum_{i=1}^n T_i Y_i/\hp_i}{\sum_{i=1}^n T_i/\hp_i} - \frac{\sum_{i=1}^n (1-T_i) Y_i/(1-\hp_i)}{\sum_{i=1}^n (1-T_i)/(1-\hp_i)}.
\end{eqnarray*}
Apparently the H\'{a}jek estimator, which normalizes the HT estimator, is a special inverse propensity score weighting estimator. 



\section{Simulation} \label{sec:simu}
In this section we present a simulation study to evaluate and compare a few propensity score estimation methods in terms of the performances of their resulting average treatment effect estimations. 

We had $1,000$ simulation runs where 
we generated independent subjects with sample size $n=200$ and $500$ respectively. For the $i$th subject, $Z_{i1}, \ldots, Z_{i6}$ were \iid sampled from the standard normal distribution. The multivariate covariate $\bW_i=(W_{i1}, W_{i2}, W_{i3})^\top$ was generated by $W_{i1} = Z_{i1} + 2Z_{i2}$, $W_{i2}=Z_{i2}^2 - Z_{i3}^2$, $W_{i3}=\exp(Z_{i3}) - \exp(1/2)$. The functional covariate was generated by $X_i(t) = \sum_{k=1}^6 A_{ik} \phi_k(t), t \in [0, 1]$ where $A_{ik}= 2 Z_{ik}/k, k=1, \ldots, 6,$ and $\phi_{2k-1}(t) = \sqrt{2}\cos(2 \pi k t)$, $\phi_{2k}(t) = \sqrt{2}\sin(2 \pi k t)$, $k=1, 2, 3$. 
Note that $E(\bW_i)=\bzero$ and $E\{X_i(t)\}=0, t\in [0, 1]$.


We generated the treatment $T_i$ using the three propensity score models (PSMs) for $p(\bW_i, X_i)$ as follows.

\begin{enumerate}
\item PSM 1: 
The treatment $T_i$ follows a Bernoulli distribution with the probability
$$
p(\bW_i, X_i) = \frac{\exp\{\alpham^\top \bW_i + \int_0^1 \beta_0(t) X_i(t)\}}{1+ \exp\{\alpham^\top \bW_i + \int_0^1 \beta_0(t) X_i(t)\}},
$$
where $\alpham = (-1, 0.5, -0.1)^\top$ and $\beta_0(t) = 2 \phi_1(t) + 0.5\phi_2(t) + 0.5\phi_3(t) + \phi_4(t)$.\\

\item PSM 2: The treatment $T_i$ follows a Bernoulli distribution with the probability
$$
p(\bW_i, X_i) = \frac{\exp\{\alpham^\top \bW_i + \int_0^1 \eta_0(t, X_i(t))\, dt\}}{1+ \exp\{\alpham^\top \bW_i + \int_0^1 \eta_0(t, X_i(t))\, dt\}},
$$
where $\alpham = (-1, 0.5, -0.1)^\top$ and $\eta_0(t, x) = -0.5 + \exp[-\{(t-0.5)/0.3\}^2 - (x/5)^2]$. \\

\item PSM 3: The treatment $T_i$ follows a Bernoulli distribution with the probability
$$
p(\bW_i, X_i) = \frac{\exp(-Z_{i1} + 0.5 Z_{i2}-0.25 Z_{i3} -0.1 Z_{i4} )}{1+ \exp(-Z_{i1} + 0.5 Z_{i2}-0.25 Z_{i3} -0.1 Z_{i4} )}.
$$
\end{enumerate}
Obviously PSM 1 is a GFPLM as in (\ref{eq:gfplm}) and PSM 2 is an FGAM as in (\ref{eq:fgam}).

%


We generated the outcome $Y_i$ based on the following two outcome models (OMs). 
\begin{enumerate}
\item OM 1: 
$Y_i = 200 + 10 T_i + (1.5 T_i - 0.5)(27.4 Z_{i1} + 13.7 Z_{i2} + 13.7 Z_{i3} + 13.7 Z_{i4}) + e_i$ where $e_i$ is generated from the standard normal distribution independently of $Z_{i1}, \ldots, Z_{i6}$. The true average treatment effect is $\tau = 10.$
\item OM 2: 
$Y_i = Z_{i1} Z_{i2}^3 Z_{i3}^2 Z_{i4} + e_i$ where $e_i$ follows the standard normal distribution which is independent of $Z_{i1}, \ldots, Z_{i6}$. The true average treatment effect is $\tau = 0.$
\end{enumerate}

We compared the performances of five propensity score estimation methods in the simulation study, denoted by \textsf{GFPLM}, \textsf{FGAM}, \textsf{CBPS1}, \textsf{CBPS2} and \textsf{KBCB} respectively. The first two methods are via direct modeling while the last three are via covariate balancing. By the FPCA approximation and maximum likelihood method as in Section \ref{sec:dm}, \textsf{GFPLM} 
fits (\ref{eq:gfplm}) to estimate the propensity score. The number of FPC scores $L$ was selected as the smallest integer such that the fraction of variation explained by the top $L$ FPC scores is at least 95\%. 
\textsf{FGAM} obtains the propensity score estimate by fitting (\ref{eq:fgam}) directly, where tensor products of seven cubic B-spline basis functions were used to approximate $\eta(*, \cdot)$ before the maximum likelihood method was applied. 
Apparently \textsf{GFPLM} is subject to model misspecification when data are generated from PSM 2, while both \textsf{GFPLM} and \textsf{FGAM} fit incorrect models when data are generated from PSM 3.

Both \textsf{CBPS1} and \textsf{CBPS2} are based on 
the CBPS method as introduced in Section \ref{sec:cb}. The multivariate substitute $\bA_i=(A_{i1}, \ldots, A_{iL})^\top$ for the functional covariate $X_i$ was obtained by FPCA which explains at least 95\% of the variation of $X_i$, and 
(\ref{eq:gfplmc}) was assumed for the substitute propensity score $p(\bC_i)$. \textsf{CBPS1} balanced the first moments of $\bC_i$ while \textsf{CBPS2} balanced both first and second moments of $\bC_i$, 
and they were performed using the \textit{CBPS} R package. \textsf{KBCB} is 
another covariate functional balancing method recently proposed by \citet{WongC18}, 
which controls the balance of $\bC_i$ over a reproducing kernel Hilbert space (RKHS). \textsf{KBCB} was implemented using the \textit{ATE.ncb} R package downloaded from \url{https://github.com/raymondkww/ATE.ncb} where the RKHS was chosen  as the second-order Sobolev space. 

\begin{table}[H]
\centering
\caption{Bias and RMSE values for the HT and H\'{a}jek estimates based on five propensity score estimation methods for PSM 1 and OM 1. The percentages beside a propensity score estimation method, if any, refer to the proportions of simulation runs used to calculate the bias and RMSE values for the HT and H\'{a}jek estimates respectively, and ``-'' denotes 100\%. All simulation runs were used for a propensity score estimation method if no such percentages are given.}
\label{tab:psm1om1}
\begin{tabular}{lcccc}
  \hline
  \hline
  \multirow{2}*{} & \multicolumn{2}{c}{HT} & \multicolumn{2}{c}{H\'{a}jek}\\
  \cmidrule(lr){2-3}\cmidrule(lr){4-5}
  & Bias & RMSE & Bias & RMSE \\ 
  \hline
  n = 200\\
 \tGFPLM (99.9\%, -) & 4.82 & 100.17 & 2.28 & 11.28 \\ 
  \tFGAM (99.9\%, -) & 8.44 & 17.12 & 9.58 & 10.47 \\
  \tCBPS1 & 2.90 & 45.87 & 3.00 & 9.15 \\ 
  \tCBPS2 (99.8\%, -)& 1.77 & 28.58 & 3.75 & 7.29 \\
  \tKBCB & 1.69 & 4.43 & 2.08 & 4.58 \\ 
  \hline
  n = 500\\
  \tGFPLM & 1.78 & 75.83 & 1.65 & 10.07 \\ 
  \tFGAM & 9.02 & 11.98 & 9.56 & 9.86 \\ 
  \tCBPS1 & 2.47 & 39.25 & 2.27 & 7.58 \\ 
    \tCBPS2 (99.8\%, -) & 1.98& 21.26 & 2.73 & 5.74 \\
  \tKBCB & 0.79 & 2.56 & 0.96 & 2.62 \\ 
   \hline
\end{tabular}
\end{table}



With the propensity score estimate obtained by each of the five methods above, 
we achieved the HT and H\'{a}jek estimates for the average treatment effect, i.e., $\htau_\text{HT}$ and $\htau_\text{H\'{a}jek}$ as in Section \ref{sec:ate}. Note that \textsf{KBCB} does not give an estimate for the substitute propensity score $p(\bC_i)$. Instead it 
provides estimates for both $T_i/p(\bC_i)$ and $(1-T_i)/\{1-p(\bC_i)\}$, but they suffice to obtain
both HT and H\'{a}jek estimates. 

\begin{table}[H]
\centering
\caption{The same as Table \ref{tab:psm1om1} except for PSM 2 and OM 1.}
\label{tab:psm2om1}
\begin{tabular}{lcccc}
  \hline
  \hline
  \multirow{2}*{} & \multicolumn{2}{c}{HT} & \multicolumn{2}{c}{H\'{a}jek}\\
  \cmidrule(lr){2-3}\cmidrule(lr){4-5}
  & Bias & RMSE & Bias & RMSE \\ 
  \hline
  n = 200\\
    \tGFPLM (99.9\%, -)& 1.26 & 57.49 & -0.68 & 8.20 \\ 
  \tFGAM (99.8\%, -) & -1.05 & 59.70 & -0.93 & 8.22 \\
  \tCBPS1 & -1.22 & 29.65 & -1.44 & 6.48 \\ 
    \tCBPS2 (99.9\%, -) & -4.19 & 22.22 & -1.81 & 5.62 \\
  \tKBCB & -1.03 & 4.11 & -0.49 & 3.95 \\ 
  \hline
  n = 500\\
  \tGFPLM & -2.46 & 40.28 & -0.87 & 6.00 \\ 
 \tFGAM & -3.23 & 43.39 & -0.74 & 5.95 \\ 
  \tCBPS1 & -0.38 & 22.52 & -1.14 & 4.94 \\ 
  \tCBPS2 & -0.98 & 15.31 & -0.98 & 3.83 \\ 
  \tKBCB & -0.35 & 2.53 & -0.17 & 2.51 \\ 
   \hline
\end{tabular}
\end{table}

The bias and root mean squared error (RMSE) values for the HT and H\'{a}jek estimates are given in Tables \ref{tab:psm1om1}--\ref{tab:psm3om2}. For each average treatment effect estimate based on any propensity score estimation method, we removed the simulation runs of which average treatment effect estimates are ten standard deviations away from the mean, and used the remaining simulated data to calculate bias and RMSE values.



\begin{table}[H]
\centering
\caption{The same as Table \ref{tab:psm1om1} except for PSM 3 and OM 1.}
\label{tab:psm3om1}
\begin{tabular}{lcccc}
  \hline
  \hline
  \multirow{2}*{} & \multicolumn{2}{c}{HT} & \multicolumn{2}{c}{H\'{a}jek}\\
  \cmidrule(lr){2-3}\cmidrule(lr){4-5}
  & Bias & RMSE & Bias & RMSE \\ 
  \hline
  n = 200\\
    \tGFPLM (99.9\%, -)& 0.15 & 22.19 & -0.15 & 5.66 \\ 
  \tFGAM (99.9\%, -) & -4.22 & 9.16 & -4.37 & 5.96 \\
  \tCBPS1 & -1.83 & 13.02 & -1.14 & 4.63 \\ 
  \tCBPS2 (99.8\%, -) & -1.69 & 14.80 & -1.11 & 4.83 \\
  \tKBCB & -0.37 & 3.92 & -0.27 & 3.90 \\ 
  \hline
  n=500\\
  \tGFPLM & -0.04 & 12.51 & -0.09 & 3.45 \\ 
  \tFGAM & -4.03 & 5.84 & -4.37 & 5.05 \\ 
  \tCBPS1 & -0.63 & 9.12 & -0.70 & 3.06 \\ 
  \tCBPS2 & -1.27 & 8.40 & -0.81 & 2.91 \\ 
  \tKBCB & -0.07 & 2.46 & -0.05 & 2.46 \\ 
   \hline
\end{tabular}
\end{table}

\begin{table}[H]
\centering
\caption{The same as Table \ref{tab:psm1om1} except for PSM 1 and OM 2. All bias and RMSE values are given in the unit of $10^{-1}$.}
\label{tab:psm1om2}
\begin{tabular}{lcccc}
  \hline
  \hline
  \multirow{2}*{$10^{-1}$} & \multicolumn{2}{c}{HT} & \multicolumn{2}{c}{H\'{a}jek}\\
  \cmidrule(lr){2-3}\cmidrule(lr){4-5}
  & Bias & RMSE & Bias & RMSE \\ 
  \hline
  n = 200\\
  \tGFPLM (99.9\%, 99.8\%) & 0.05 & 18.26 & 0.27 & 12.48 \\ 
  \tFGAM (99.9\%, 99.9\%)& 3.28 & 11.83 & 3.29 & 11.75 \\
  \tCBPS1 (99.9\%, 99.9\%) & 0.59 &9.74 & 0.65 & 10.71 \\
  \tCBPS2 (99.8\%, -)& 0.99& 7.92 & 1.04 & 8.26 \\
  \tKBCB & 0.95 & 8.70 & 0.95 & 8.71 \\
  \hline
  n = 500\\ 
  \tGFPLM (99.8\%, 99.9\%) & -0.07 & 11.59 & -0.04 & 10.76 \\
  \tFGAM (99.8\%, 99.8\%) & 4.01 & 10.40 & 4.02 & 10.43 \\
  \tCBPS1 (99.9\%, 99.9\%)& 0.11 & 8.49 & 0.22 & 8.63 \\
  \tCBPS2 (99.8\%, 99.9\%) & 0.51 & 6.74 & 0.64 & 6.58 \\
  \tKBCB & 1.04 & 6.71 & 1.04 & 6.71 \\ 
   \hline
\end{tabular}
\end{table}




The six tables show that for any propensity score estimation methods but \textsf{KBCB}, 
a larger sample size generally improves the average treatment effect estimation accuracy 
measured by RMSE, 
but it unnecessarily improves the bias. With respect to RMSE, the three covariate balancing methods are generally better than the two directly modeling methods, although \textsf{FGAM} occasionally outperforms the two \textsf{CBPS} methods (see Tables \ref{tab:psm1om1} and \ref{tab:psm3om1}).
Between the two direct modeling methods, \textsf{FGAM} almost always performs better than \textsf{GFPLM} in terms of RMSE even when the latter correctly specifies PSM 1, but the former is often worse in terms of bias. 
The results for the two \textsf{CBPS} methods indicate that balancing additional covariate moments can typically improve average treatment effect estimation. Among all five propensity score estimation methods, \textsf{KBCB} performs overall the best in terms of both bias and RMSE with the only exceptions for PSM 1 and OM 2 (see Table \ref{tab:psm1om2}) and for PSM 2 and OM 2 with $n=200$ (see Table \ref{tab:psm2om2}).

\begin{table}[H]
\centering
\caption{The same as Table \ref{tab:psm1om2} except for PSM 2 and OM 2.}
\label{tab:psm2om2}
\begin{tabular}{lcccc}
  \hline
  \hline
  \multirow{2}*{$10^{-1}$} & \multicolumn{2}{c}{HT} & \multicolumn{2}{c}{H\'{a}jek}\\
  \cmidrule(lr){2-3}\cmidrule(lr){4-5}
  & Bias & RMSE & Bias & RMSE \\ 
  \hline
  n = 200\\
  \tGFPLM (99.8\%, 99.8\%)& -0.54 & 10.24 & -0.52 & 9.29 \\
  \tFGAM (99.7\%, 99.9\%) & -0.21 & 10.17 & -0.24 & 9.91 \\
  \tCBPS1 & -0.50 & 8.43 & -0.51 & 8.87 \\ 
  \tCBPS2 (99.9\%, -)& -0.35 & 5.95 & -0.34 & 6.70 \\
  \tKBCB & -0.66 & 7.13 & -0.66 & 7.14 \\ 
  \hline
  n = 500\\
  \tGFPLM (99.8\%, 99.9\%)& -0.53 & 8.46 & -0.54 & 8.25 \\
  \tFGAM (99.8\%, 99.9\%)& -0.54 & 9.70 & -0.55 & 8.96 \\
  \tCBPS1 & -0.62 & 6.86 & -0.67 & 7.17 \\ 
  \tCBPS2 (99.9\%, 99.9\%)& -0.55 & 6.24 & -0.60 & 6.35 \\
  \tKBCB & -0.60 & 5.12 & -0.60 & 5.12 \\ 
   \hline
\end{tabular}
\end{table}




\begin{table}[H]
\centering
\caption{The same as Table \ref{tab:psm1om2} except for PSM 3 and OM 2.}
\label{tab:psm3om2}
\begin{tabular}{lcccc}
  \hline
  \hline
  \multirow{2}*{$10^{-1}$} & \multicolumn{2}{c}{HT} & \multicolumn{2}{c}{H\'{a}jek}\\
  \cmidrule(lr){2-3}\cmidrule(lr){4-5}
  & Bias & RMSE & Bias & RMSE \\ 
  \hline
  n = 200\\
  \tGFPLM (99.9\%, 99.9\%)& -0.43 & 10.91 & -0.39 & 10.49 \\
  \tFGAM (99.8\%, 99.8\%) & -0.29 & 10.19 & -0.30 & 10.31\\
  \tCBPS1 & -0.35 & 8.43 & -0.37 & 8.75 \\ 
  \tCBPS2 (99.9\%, 99.9\%) & -0.37 & 10.43 & -0.31 & 8.94 \\
  \tKBCB & -0.12 & 7.10 & -0.12 & 7.11 \\ 
  \hline
  n = 500\\
  \tGFPLM & -0.92 & 7.44 & -0.90 & 7.35 \\ 
  \tFGAM & -1.12 & 7.91 & -1.13 & 7.92 \\ 
  \tCBPS1 & -0.70 & 6.29 & -0.72 & 6.42 \\ 
  \tCBPS2 (99.9\%, 99.9\%) & -0.59 & 5.89 & -0.60 & 6.04 \\
  \tKBCB & -0.30 & 4.88 & -0.30 & 4.88 \\ 
   \hline
\end{tabular}
\end{table}

%

In terms of computational stability, all propensity score estimation methods perform satisfactorily, but \textsf{KBCB} is the most robust method since it never produces outlying average treatment effect estimates. \textsf{CBPS}1 is slightly less likely to produce extreme average treatment effect estimates than 
\textsf{CBPS}2. This is somewhat unsurprising 
since the latter additionally balances the second moments of covariates. Compare to the HT estimates, the H\'{a}jek estimates generally have fewer outlying values and smaller 
RMSE values for all propensity score estimation methods but \textsf{KBCB}. This observation demonstrates the benefit of inverse propensity score weighting.

\section{Data Application} \label{sec:data}


We applied three propensity score weighting methods introduced above to a pain relief dataset \citep{TetrMV16}, 
which was downloaded from OpenNeuro
(\url{https://openneuro.org/datasets/ds000208/versions/1.0.0}).
The dataset consists of $56$ chronic knee osteoarthritis pain patients in two separate clinical trials. The first trial was single-blind where all $17$ subjects took placebo pills, while the second trial was double-blind
where $39$ subjects were randomized to take either duloxetine (30/60mg QD) or placebo. 
With the observational data obtained by combining the two trials,
we aimed to estimate the average treatment effect of duloxitine compared to placebos on chronic knee osteoarthritis pain relief. The pain relief was measured by the visual
analog scale (VAS) score, and the Western Ontario and McMaster Universities Osteoarthritis
Index (WOMAC) score, and we studied the average duloxitine effect on both measures separately. 

\begin{table}[H] 
\caption{
The HT and H\'{a}jek estimates for the average treatment effect of duloxitine on pain relief measured by the VAS score. 
Bootstrap standard errors (SE) and 95\% bootstrap percentile confidence intervals were obtained by 1,000 bootstrap samples.}
\label{tab:VAS}
\centering
\begin{tabular}{lcccc}
  \hline
  \hline
 & $\htau$ & $SE$ & [2.5\% , 97.5\%] \\ 
  \hline
  HT & & &\\
  \tGFPLM & -5.99 & 11.62 & [-22.77 , 6.49] \\ 
  \tCBPS & -5.26 & 4.12 & [-12.21 , 3.17] \\ 
  \tKBCB & 0.39 & 3.43 & [-6.29 , 7.21] \\ 
  \hline
  H\'{a}jek & & &\\
  \tGFPLM & -0.52 & 4.27 & [-9.04 , 7.58] \\ 
  \tCBPS & -0.23 & 3.64 & [-6.87 , 7.18] \\ 
  \tKBCB & 0.28 & 3.37 & [-6.14 , 7.17]\\ 
   \hline
\end{tabular}
\end{table}


A subject is considered to receive the treatment if he/she took duloxitine; 
those who took placebo pills are regarded to be in the control group. 
The multivariate covariates $\bW_i$ are age and gender. 
Each subject also underwent pretreatment brain scans, via both anatomical magnetic
resonance imaging (MRI) and resting state functional MRI (rsfMRI). Using the
FMRIB Software Library v6.0 (\url{https://fsl.fmrib.ox.ac.uk/fsl/fslwiki}),
we preprocessed the rsfMRI scans of each subject, 
registered them to template MNI152 through his/her anatomical MRI scan, and then downsampled each registered rsfMRI scan to the spatial resolution of voxel size $4\text{mm}^3$. 
Finally, inspired by \cite{TetrMV16}, we obtained a connectivity network/matrix for each subject which contains 
the Pearson correlation of the brain signals from every pair of voxels, and we treated this network as a functional covariate $X_i$.
Since each voxel is indexed by a three-dimensional spatial coordinate, the functional covariate is six-dimensional. 


\begin{table}[H] 
\caption{The same as Table \ref{tab:VAS} except for the WOMAC score.}
\label{tab:WOMAC}
\centering
\begin{tabular}{lcccc}
  \hline
  \hline
 & $\htau$ & $SE$ & [2.5\% , 97.5\%] \\ 
  \hline
  HT & & &\\
  \tGFPLM & -8.41 & 11.23 & [-26.25 , 5.35] \\ 
  \tCBPS & -7.63 & 4.79 & [-16.22 , 2.08] \\ 
  \tKBCB & 1.16 & 4.18 & [-7.54 , 8.45] \\ 
  \hline
  H\'{a}jek & & &\\
  \tGFPLM & -3.18 & 4.72 & [-11.66 , 6.50] \\ 
  \tCBPS & -2.85 & 4.31 & [-10.56 , 6.52] \\ 
  \tKBCB & 1.05 & 4.17 & [-7.76 , 8.50] \\ 
   \hline
\end{tabular}
\end{table}

We considered three methods for propensity score estimation, \textsf{GFPLM}, 
\textsf{CBPS} and \textsf{KBCB}. 
\textsf{GFPLM} refers to the direct modeling method where the propensity score is estimated by fitting the model in Remark \ref{rmk:gfplm}.2.
The top $L$ FPC scores of $X_i$, denoted by 
$\bA_i=(A_{i1}, \ldots, A_{iL})^\top$, 
were used in the approximate model 
(\ref{eq:agfplm}) for \textsf{GFPLM}. They were also used as the multivariate substitute of $X_i$ 
to define the substitute propensity score as in (\ref{eq:aps}) 
to perform \textsf{CBPS} and \textsf{KBCB}. To apply \textsf{CBPS}, (\ref{eq:gfplmc}) was assumed for the substitute propensity score, and only the first moments of $\bC_i = (\bW_i^\top, \bA_i^\top)^\top$ were balanced due to a small sample size.  
We used $L=4$ in all three methods, which was selected as the smallest integer 
such that the corresponding AIC value no longer decreases
when the top FPC scores are sequentially added to (\ref{eq:agfplm}). 

For each propensity score estimation and each pain relief measure, i.e., VAS or WOMAC score, we obtained its corresponding HT and H\'{a}jek estimates for the average treatment effect of duloxitine.
We used $1,000$ bootstrap samples to provide uncertainty measures, 
including standard errors and confidence intervals.

\begin{figure}[H] 
  \centering
  \includegraphics[width=9cm]{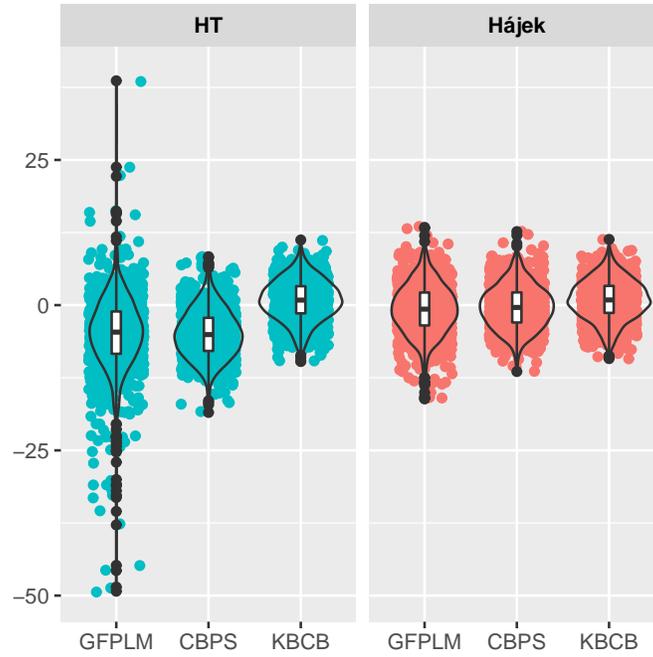}
  \caption{Violin plots for the 1,000 bootstrap HT and H\'{a}jek estimates for the average treatment effect of duloxitine on the VAS score.}
  \label{fig:VAS}
\end{figure}

\begin{figure}[H] 
  \centering
  \includegraphics[width=9cm]{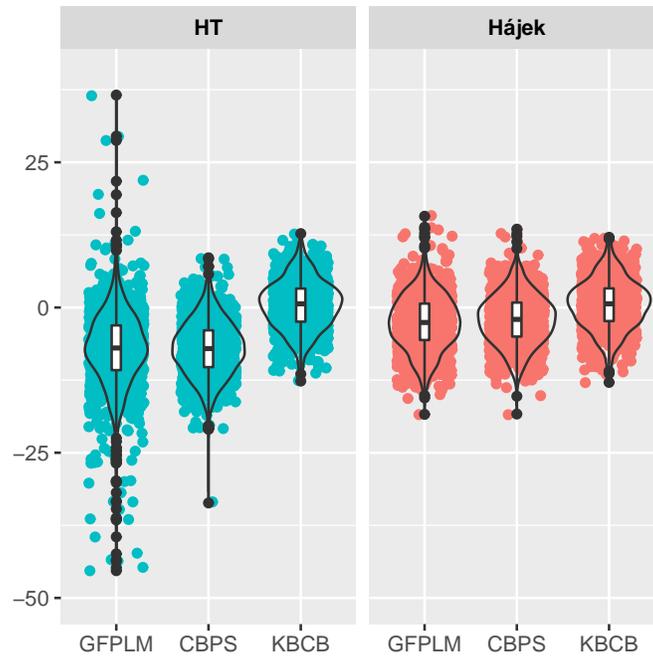}
  \caption{The same as Figure \ref{fig:VAS} except for the WOMAC score.}
  \label{fig:WOMAC}
\end{figure}

Tables \ref{tab:VAS} and \ref{tab:WOMAC} provide the HT and H\'{a}jek estimates, bootstrap standard errors and 95\% bootstrap percentile confidence intervals for the average treatment effect of duloxitine on VAS and WOMAC scores respectively. 
They indicate no significant treatment effect of duloxitine over placebo pills on pain relief.
This is consistent with 
\citet{TetrMV16}, although their conclusion was made from
a double-blind clinical trial, i.e., the second trial, 
while we based our finding on an observational dataset.
{Explicitly, the 95\% confidence intervals for the average treatment effect of duloxitine on VAS and WOMAC scores obtained from the double-blind trial are $[-8.402, 6.760]$ and $[-9.717, 8.832]$ respectively. Both confidence intervals are based on t statistics assuming 
normality of each treatment or control group and equal variances of the two groups, 
which are validated by the Shapiro-Wilk tests and F-tests respectively.}


The $1,000$ bootstrap HT and H\'{a}jek average treatment effect 
estimates are illustrated in Figures \ref{fig:VAS} and \ref{fig:WOMAC} for VAS and WOMAC scores respectively. 
Both figures show that the HT estimates based on 
\textsf{GFPLM} for propensity score estimation have a much larger variation than the two covariate balancing methods, but inverse propensity score weighting can substantially reduce their differences as revealed by the H\'{a}jek estimates. 
The median of the H\'{a}jek estimates is shifted towards zero compared to that of the HT estimates when the propensity score is estimated by either \textsf{GFPLM} or \textsf{CBPS}. The two average treatment effect estimates essentially show no difference for \textsf{KBCB}. 





\section{Discussion} \label{sec:disc}

To the best of our knowledge, this paper has made the first attempt to study average treatment effect estimation via propensity score weighting for functional data in observational studies. The paper introduces 
both direct modeling and covariate balancing methods for 
propensity score estimation and systematically evaluates their performances via a simulation experiment and a real data application. The results confirm the benefits of both inverse propensity score weighting and covariate balancing methods as  advocated for multivariate data.

The methods introduced in this paper for average treatment effect estimation only focus on the scenario where the outcome is a continuous scalar variable and there is only one functional covariate. However, with straightforward modifications, they may be generalized to handle multiple functional covariates and continuous functional outcomes.

The covariate balancing methods introduced above rely on a satisfactory multivariate substitute for the functional covariate, which requires the functional covariate to be either fully observed or densely measured \citep[e.g.,][]{DauxPR82, HallH06, HallMW06}. A future research topic is to develop covariate balancing methods 
for sparsely measured functional covariates \citep[e.g,][]{JameHS00, YaoMW05a} or a unified approach for all types of functional covariates \citep[e.g.,][]{LiH10, ZhanW16, Lieb19}. {Another interesting direction is to study non-truncation regularization methods, e.g., the roughness penalization, to solve covariate balancing equations with functional covariates.}



\section*{Acknowledgements}\label{sec:ack}
Xiaoke Zhang's research was partly supported by the USA National Science Foundation under grant DMS-1832046. 


\bibliographystyle{chicago}
\bibliography{mfar,nsf-xiaoke}          
%
%

\end{document}